\documentclass[a4paper,11pt]{article}
\pdfoutput=1
\usepackage[colorlinks=true]{hyperref}
\usepackage{graphicx,rotating,hyperref,slashed,amsmath,xcolor,amssymb,amsfonts,colortbl,cite}
\makeatletter
\hypersetup{colorlinks,bookmarksopen,bookmarksnumbered,
linkcolor=blus,pdfstartview=FitH,urlcolor=rossos,citecolor=verde}
\allowdisplaybreaks		
\hypersetup{%
    ,urlcolor=black
    ,citecolor=black
    ,linkcolor=black
    }
\usepackage{mciteplus} 

\AtBeginDocument{
  \hypersetup{
    urlcolor=black,
    citecolor=black,
    linkcolor=blue,
  }%
}

\numberwithin{equation}{section}
 
\def\beq{\begin{equation}}
\def\eeq{\end{equation}}

\def\bea{\begin{eqnarray}}
\def\eea{\end{eqnarray}}

\def\0{{\boldsymbol 0}}

\def\lsim{\mathrel{\rlap{\lower3pt\hbox{\hskip0pt$\sim$}}
   \raise1pt\hbox{$<$}}}         
\def\gsim{\mathrel{\rlap{\lower4pt\hbox{\hskip1pt$\sim$}}
   \raise1pt\hbox{$>$}}}         

 \newcommand{\sfootnote}[1]{}
\definecolor{bluc}{cmyk}{1,1,0,0.1}
\definecolor{rossoCP3}{cmyk}{0,.88,.77,.40}
\definecolor{rosso}{cmyk}{0,1,1,0.4}
\definecolor{rossos}{cmyk}{0,1,1,0.55}
\definecolor{rossoc}{cmyk}{0,1,1,0.2}
\definecolor{verdes}{cmyk}{0.92,0,0.59,0.4}

\hypersetup{colorlinks, bookmarksopen, bookmarksnumbered,
citecolor=verdes, linkcolor=bluc, pdfstartview=FitH, urlcolor=rossos}

\newcommand{\mio}[1]{}

\definecolor{Gray}{gray}{0.95}

\usepackage{multicol}
\usepackage{color}
\definecolor{rosso}{cmyk}{0,1,1,0.4}
\definecolor{rossos}{cmyk}{0,1,1,0.55}
\definecolor{rossoc}{cmyk}{0,1,1,0.2}
\definecolor{blu}{cmyk}{1,1,0,0.3}
\definecolor{blus}{cmyk}{1,1,0,0.6}
\definecolor{bluc}{cmyk}{1,1,0,0.1}
\definecolor{verde}{cmyk}{0.92,0,0.59,0.25}
\definecolor{verdec}{cmyk}{0.92,0,0.59,0.15}
\definecolor{verdes}{cmyk}{0.92,0,0.59,0.4}

\skip\@mpfootins = \skip\footins
\renewcommand\&{&}

\def\circa#1{\,\raise.3ex\hbox{$#1$\kern-.75em\lower1ex\hbox{$\sim$}}\,}

\newcommand{\be}{\begin{equation}}
\newcommand{\ee}{\end{equation}}
\newfam\rsfsfam
\def\mathscr#1{{\fam\rsfsfam\relax#1}}

\def\circa#1{\,\raise.3ex\hbox{$#1$\kern-.75em\lower1ex\hbox{$\sim$}}\,}
\makeatletter

\def\hhref#1{\href{http://arxiv.org/abs/#1}{arXiv:#1}} 

\newcommand{\doi}[1]{\href{http://dx.doi.org/#1}{[doi]}}

\def\hhref#1{\href{http://arxiv.org/abs/#1}{arXiv:#1}} 
 
\def\art{\@ifnextchar[{\eart}{\oart}}
\def\eart[#1]#2#3#4#5#6{{\rm #2}, {\em #3 \bf #4} {\rm (#6) #5} ({\em #1})}

\def\article{\@ifnextchar[{\earticle}{\oarticle}}
\def\oarticle#1#2#3#4#5#6{{\rm #1}, {\em ``#6''}, {\rm #2 #3 (#5) #4}}
\def\earticle[#1]#2#3#4#5#6#7{{\rm #2}, {\em ``#7''}, {\rm #3 #4 (#6) #5}  [\hhref{#1}]}
\def\hepart[#1]#2{{\rm #2, \em#1}}
\def\heparticle[#1]#2#3{#2, {\em ``#3''} [\hhref{#1}]}

\newcounter{alphaequation}[equation]
\def\thealphaequation{\theequation\hbox to
0.6em{\hfil\alph{alphaequation}\hfil}}
\def\eqnsystem#1{
\def\@eqnnum{{\rm (\thealphaequation)}}
\def\@@eqncr{\let\@tempa\relax \ifcase\@eqcnt \def\@tempa{& & &} \or
  \def\@tempa{& &}\or \def\@tempa{&}\fi\@tempa
  \if@eqnsw\@eqnnum\refstepcounter{alphaequation}\fi
\global\@eqnswtrue\global\@eqcnt=0\cr}
\refstepcounter{equation} \let\@currentlabel\theequation \def\@tempb{#1}
\ifx\@tempb\empty\else\label{#1}\fi
\refstepcounter{alphaequation}
\let\@currentlabel\thealphaequation
\global\@eqnswtrue\global\@eqcnt=0 \tabskip\@centering\let\\=\@eqncr
$$\halign to \displaywidth\bgroup \@eqnsel\hskip\@centering
$\displaystyle\tabskip\z@{##}$&\global\@eqcnt\@ne
\hskip2\arraycolsep\hfil${##}$\hfil& \global\@eqcnt\tw@\hskip2\arraycolsep
$\displaystyle\tabskip\z@{##}$\hfil
\tabskip\@centering&\llap{##}\tabskip\z@\cr}
\def\endeqnsystem{\@@eqncr\egroup$$\global\@ignoretrue} \makeatother

\definecolor{fiorentina}{rgb}{.5,0,.5}

\begin{document}

\setcounter{page}{1} \baselineskip=15.5pt \thispagestyle{empty}

\bigskip\

\vspace{1cm}
\begin{center}

{\fontsize{19}{28}\selectfont  \sffamily \bfseries {Symmetries for scalarless scalar theories}}

\end{center}

\vspace{0.2cm}

\begin{center}
{\fontsize{13}{30}\selectfont Gianmassimo Tasinato} 
\end{center}

\begin{center}

\vskip 8pt
\textsl{ Physics Department, Swansea University, SA28PP, UK}\\
\vskip 7pt

\end{center}

\smallskip
\begin{abstract}
\noindent
We consider  theories containing   scalar fields interacting with vector or with tensor degrees of freedom,  equipped with  symmetries that prevent the propagation of linearized scalar excitations around solutions of the equations of motion.  We first study  the implications of such symmetries for building vector theories   that break Abelian gauge invariance without   necessarily exciting longitudinal scalar fluctuations  in flat space.  We then examine scalar-tensor theories in curved space, and  relate the symmetries we consider with a non-linear realization of broken  space-time symmetries acting on scalar modes. We  determine sufficient conditions on the space-time geometry  to avoid the propagation of scalar fluctuations. We analyse linearized perturbations  around spherically symmetric black holes,  proving   the absence of scalar excitations, and pointing out  modifications  in the dynamics of spin-2 fluctuations with respect to Einstein gravity. We then study  consequences of  this set-up for   the dark energy problem,  determining  scalar constraints on cosmological configurations that can  lead to  self-accelerating universes whose expansion is  insensitive to the value of the  bare cosmological constant. 
\end{abstract}

\section{Introduction}

In this work we examine covariant theories describing scalar fields interacting with themselves, with vectors, or
with tensors, where  symmetries prevent the propagation
of  linearized scalar excitations around solutions of the equations of motion.  
  Our motivations for studying these systems are the following:
 \begin{itemize}
\item[-]  They  can constitute examples of theories of gravity alternative
 to General Relativity (GR) that
 automatically avoid fifth force constraints, since
 they only admit  a massless spin-2 propagating mode. In fact, if scalar fluctuations do not
 propagate, long range scalar interactions
 are absent. 
  These systems
   circumvent Lovelock theorem \cite{Lovelock:1971yv} by
 spontaneously breaking Lorentz invariance by a non-trivial profile
 for the scalar 
 background configuration.
 \item[-] These theories might  avoid instabilities associated with scalar fluctuations in scalar-tensor systems,
 as for example
graviton
 decay into dark energy \cite{Creminelli:2018xsv},
   or scalar instabilities around spherically symmetric black holes
\cite{Ogawa_2016,Takahashi_2017,Khoury:2020aya,Babichev_2018,Minamitsuji:2018vuw,de_Rham_2019,Motohashi:2019ymr}. Yet, the dynamics
of tensor fluctuations can be distinct with respect to GR, making them distinguishable from Einstein gravity.
 \item[-]  Symmetries  can  limit the number of allowed interactions for the set-up
 under examination, and protect their structure under classical
   and quantum corrections. The structure of the corresponding theories can lead to new perspectives
   for the dark energy problem, and for explaining the smallness of the present-day cosmological acceleration rate
   \cite{Weinberg:1988cp}. 
\end{itemize}

 \smallskip

 Examples of scalar-tensor or massive gravity theories where scalar fluctuations do not propagate have
 already been discussed in the literature, see e.g. \cite{Lin_2017,Aoki_2018,Afshordi_2007,Iyonaga_2018,Gao_2020}, usually performing   a Hamiltonian analysis
  aimed at determining the constraints that avoid the propagation of scalar modes. In many of the existing
  examples, scalar fluctuations do not propagate in a unitary gauge, where   an homogeneous, time-dependent-only configuration for the background scalar field is selected. In this work, we address
  the problem of determining scalarless scalar-tensor theories of gravity making use of symmetry 
  principles, that guide us  for building covariant scalarless theories and for better understanding the dynamics
  of the propagating modes.
   We focus on  symmetries first explored in \cite{Chagoya:2016inc,Chagoya:2018yna}, that arise when
  taking certain limits of DBI-Galileon interactions \cite{deRham:2010eu}, and further develop and expand previous results. We
   aim to better understand how such scalarless interactions arise when breaking gauge or space-time symmetries
   in vector or tensor theories, with the scalar playing the role of Goldstone bosons of such symmetries.
  We then study the consequences of our results for spherically symmetric configurations and for cosmology.  
   We proceed as follows:
  \begin{itemize}
  \item We start in section \ref{symflat} considering scalar theories in flat space in absence of gravity, and studying a field-dependent
  coordinate reparametrization ($\pi(x)$ is the scalar field, $\omega^\mu$ an arbitrary infinitesimal constant
  vector)
  \be \label{introsymf}
  x^\mu\,\to\,x^\mu+ \pi(x)\,\omega^\mu\,.
  \ee
As  we shall see,  this symmetry forbids the propagation of scalar fluctuations. Scalar
     theories invariant under this transformation  inevitably lead to spontaneous breaking of Lorentz
     invariance, by turning on a non-vanishing scalar gradient $\partial_\mu \pi$.  
     The vacuum of these theories selects a preferred direction that can be identified with the
      time coordinate, making this set-up reminiscent of Einstein-Ether systems (see e.g. \cite{Jacobson:2008aj} for
      a review). 
      In our case, after Lorentz  symmetry is spontaneously broken, no scalar excitations  propagate. This
      result is not in contrast with Lovelock theorem, where Lorentz invariance is assumed. 
      \item In section \ref{sec-ab} we couple scalar modes with vector fields in flat space, showing how
      to select interactions  that -- although we break the standard Abelian $U(1)$ symmetry --
       preserve a scalar-vector generalization of symmetry \eqref{introsymf}. The system 
      spontaneously break Lorentz invariance, and the new symmetry can prevent the propagation
      of scalar excitations, leading to theories where only transverse vector modes are dynamical, 
      even in absence of Abelian $U(1)$ gauge invariance.
      \item In section \ref{sec-gra} 
      we study 4-dimensional scalar-tensor interactions in curved space, deriving them from
      a brane-world construction of a dynamical brane embedded in 5-dimensional (5d) space \cite{deRham:2010eu}. 
      We start in section \ref{sec-bw-emb} turning gravity off,  reviewing the arguments of \cite{deRham:2010eu,Chagoya:2016inc}, and showing how symmetry  
       \eqref{introsymf} arises as a non-linear realization of broken global symmetries in the 5d
       brane-world model. In section \ref{sec-gr} we then consider the case of dynamical gravity, showing that although
       symmetry  \eqref{introsymf}  gets generally broken, nevertheless a novel continuous symmetry arises,
       inherited from 5d diffeomorphism invariance. In section \ref{sec-conse} we study consequences of this symmetry,
       providing sufficient conditions for avoiding the propagation of scalar modes around
       certain configurations. As representative example,
       we demonstrate that no scalar excitations arise around a Schwarzschild solution, although
       the dynamics of spin-2 fluctuations is different with respect to Einstein gravity, making the theory
       distinguishable from General Relativity (GR). In section \ref{cc-sect} we study cosmology, pointing out
       that the structure of our theories  leads to scalar constraints on cosmological configurations, 
       leading to expanding space-time geometries insensitive to the value of the cosmological constant. Correspondingly, we 
       discuss some consequences for the dark energy problem.  
  \item We conclude our discussion in section \ref{sec-out}, followed by a technical appendix \ref{app-BHpert}. 
  \end{itemize}

 \section{A symmetry for scalarless scalar theories around flat space} \label{symflat}

\subsection{The scalarless symmetry}\label{sec-flat-sym}
 
  We start  considering a class of  scalar theories around flat space in absence of gravity, whose Langrangian densities are Lorentz invariant, and lead to  second order equations of motion for the scalar field $\pi(x)$. Besides a shift symmetry, we demand that these  theories  are  invariant under the  field-dependent coordinate transformation
  \be\label{dessymco}
  x^\mu\to x^\mu+\pi(x)\,\omega^\mu
  \ee
  with  $\omega^\mu$ a
  {\it constant, infinitesimal 4-vector}. This infinitesimal coordinate transformation acts on the scalar field $\pi(x)$ as
  \bea\label{dessym}
    \pi(x) \to \pi(x)+
  \delta_{\rm sl} \pi(x) \hskip0.5cm,\hskip0.5cm {\rm with} \hskip1cm 
   \delta_{\rm sl} \pi(x)\,\equiv\,
  \frac12\,\omega^\mu \,\partial_\mu \pi^2(x)
\,\,.  \eea 
 As we shall demonstrate soon, theories invariant under the symmetry \eqref{dessym}
 spontaneously break Lorentz symmetry, and the corresponding equations of motion
 require to turn on a non-vanishing time-like
 background gradient $\partial_\mu \bar \pi$  whose norm satisfies the condition
   $\partial_\mu  \bar \pi\, \partial^\mu \bar \pi\,<\,0$  (we work with  `mostly plus' metric
   conventions).

    On the other hand, the symmetry \eqref{dessym}  
  prevents  the  propagation of linearized scalar excitations $\hat \pi$ around 
  the   
     background configuration $\bar \pi$. 
               For this reason  we dub transformation \eqref{dessym} {\it scalarless} (SL) symmetry.  
      This fact was  first noticed in \cite{Chagoya:2016inc}, and
    we now reconsider it here. 
         We  select  a background scalar field  $\bar \pi(t)$ that spontaneously 
      break Lorentz symmetry and 
      solves
    the equations of motion.
  Splitting the scalar into a background homogenous configuration plus a small perturbation, $\pi(t,\,\vec x)\,=\,\bar \pi(t)+\hat \pi(t,\,\vec x)$,  these two quantities 
  transform under symmetry  \eqref{dessym}  as
  \bea\label{reds1}
    \bar \pi(t)& \to& \bar \pi(t)+\omega^0 \,
  \bar \pi(t)\,\dot{\bar\pi}(t)\,,
  \\ \label{reds2}
  \hat \pi(t,\,\vec x)&\to&  \hat \pi(t,\,\vec x)+\omega^0\,\partial_0 \left[
    \bar \pi(t)\, \hat \pi(t,\,\vec x)\right]+    \bar \pi(t)\,\omega^i\,\partial_i\,\hat \pi(t,\,\vec x)\,,
  \eea
  for infinitesimal quantities $\omega^\mu$ and $\hat \pi$. This implies that, after
  fixing once for all the background configuration $\bar \pi(t)$,
   there remains a residual symmetry depending on the spatial components $\omega^i$ of the vector $\omega^\mu$,  and acting on the small fluctuations as
   \be\label{dessym2}
   \hat \pi(x^\mu)\,\to\, \hat \pi(x^\mu)+ \bar \pi (t) \,\omega^i \,\partial_i \hat \pi(x^\mu)\,.
       \ee
 Such residual symmetry prevents a standard kinetic term for the fluctuations around Minkowski
 space: the  contribution $(1/2)\left(\partial_0 \,\hat \pi\right)^2$ to the scalar
 kinetic term  containing time derivatives~\footnote{
 Instead, the contribution to the kinetic term containing spatial derivatives, $(\vec \nabla \pi)^2$, is invariant under
 the  infinitesimal residual symmetry
 up to total derivatives, since
 \bea
\frac12\,\partial_j \hat \pi \partial^j \hat \pi\,\to\,\partial_j \hat \pi \partial^j \left[ \omega^i \partial_i \left( \bar \pi(t)\,\hat \pi \right)\right]\,=\,\frac{ \omega^i }{2}\,\partial_i \left[ \bar \pi(t)\,\partial_j \hat \pi \partial^j \hat \pi \right]\,.
 \eea
We will explore in the next sections some consequences of this fact. 
 } is {\it not invariant} under the residual symmetry 
 \eqref{dessym2}, hence it can not appear in the quadratic action for scalar fluctuations.

 \bigskip

       For these reasons, we regard scalar  theories invariant under   the SL symmetry   \eqref{dessym}  as a flat-space proxy for scalarless scalar-tensor
 systems  in a unitary gauge, where the scalar profile is time-dependent. For the rest of this section
 we characterise the structure and properties these scalar theories, to then explore 
 their possible physical origin  in the next sections.

 \subsection{The scalarless  scalar Lagrangians}

 Are there shift-symmetric scalar theories invariant under the infinitesimal transformation \eqref{dessym}? The answer is affirmative, and 
 these theories arise
 as a  limit of a certain generalization of  DBI-Galileons \cite{deRham:2010eu}: this fact was first shown in \cite{Chagoya:2016inc}. Here 
 we streamline and elaborate on the analysis of \cite{Chagoya:2016inc}.

 \smallskip
 
 To express more concisely the scalar  systems we are interested in, we  adopt the
 notation  $\partial_\mu \pi\,=\,\pi_{,\mu}$ for denoting derivatives. In this section we raise
and lower indexes with the Minkowski metric $\eta_{\mu\nu}$.  We define
 \be
 X\,=\,-\pi_{, \mu} \pi^{, \mu} 
 \ee
 and the convenient combinations
 \be
 [\Pi^n]\,=\,\pi^{,\mu}_{\,\,\,, \alpha_1}
\,\pi^{,\alpha_1}_{\,\,\,, \alpha_2}\,
\dots \pi^{,\alpha_{n}}_{\,\,\,,\mu} \hskip1cm,\hskip1cm
[\Phi^n]\,=\,
\pi_{, \mu}
\,\pi^{,\mu}_{\,\,\,, \alpha_1}\,
\dots \pi^{,\alpha_{n}}_{\,\,\,,\mu}\,\pi^{,\mu}\,.
 \ee
  The shift-symmetric  Lagrangian
 densities
  describing the theories respecting the symmetry \eqref{dessym}, up to total derivatives, are
  \bea
  \label{L1fl}
  {\cal L}_1&=&\sqrt{X}\,,
  \\
    \label{L2fl}
    {\cal L}_2&=&\frac{1}{X}\,[\Phi]\,,
     \\
         \label{L3fl}
    {\cal L}_3&=&\frac{1}{\sqrt{X}}\,\left([\Pi^2]-[\Pi]^2 \right)\,,
     \\
         \label{L4fl}
    {\cal L}_4&=&\frac{1}{{X}}\,\left([\Pi]^3+2\, [\Pi^3] -3 \,[\Pi^2] [\Pi]\right)\,.
  \eea
The Lagrangian \eqref{L1fl} corresponds to a flat-space version of the cuscuton
system \cite{Afshordi_2007}, while Lagrangians \eqref{L2fl}-\eqref{L4fl} might be thought
as  higher derivative versions of this system (some common properties of all these Lagrangians will be discussed
in what follows).  
These Lagrangians are weighted by appropriate powers of an energy scale (which we set to one) to provide the correct dimensionality. 

We now present a more compact way to express and analyse these Lagrangians, which allows one to drive closer parallelisms with
generalized Galileons, and
 more easily understand their symmetry properties. The previous four quantities -- up to total
derivatives and overall constants -- can be expressed as:
\be\label{LagN}
{\cal L}_{n}\,=\,\sqrt{X}\,Y_\mu^{{\bf (n)}\,\,\,\mu}\,,
\ee
for $n\,=\,1,\dots,4$, 
 where the tensors  $Y_\mu^{{\bf (n)}\,\,\,\mu}$ read
 \bea
Y^{{\bf (1)}\,\,\nu}_{\,\mu}&=&\epsilon_{\mu\, \alpha_1\beta_1\delta_1}\,\epsilon^{\nu\, \alpha_1\beta_1\delta_1}
\,\,,
\\
Y^{{\bf (2)}\,\,\nu}_{\,\mu}&=&\epsilon_{\mu\, \alpha_1\beta_1\delta_1}\,\epsilon^{\nu\, \alpha_1\beta_1\delta_2}
\,\left({\pi^{,\delta_1}}/\sqrt{X}\right)_{, \delta_2}
\,\,,
\\
Y^{{\bf (3)}\,\,\nu}_{\,\mu}&=&\epsilon_{\mu\, \alpha_1\beta_1\delta_1}\,\epsilon^{\nu\, \alpha_1\beta_2\delta_2}
\, \left(  \pi^{,\beta_1}
/\sqrt{X}
\right)_{,{\beta_2}}
 \,\left(  \pi^{,\delta_1}
/\sqrt{X}
\right)_{, \delta_2}
\,\,,
\\
Y^{{\bf (4)}\,\,\nu}_{\,\mu}&=&\epsilon_{\mu\, \alpha_1\beta_1\delta_1}\,\epsilon^{\nu\, \alpha_2\beta_2\delta_2}
\,
\, \left(  \pi^{,\alpha_1}
/\sqrt{X}
\right)_{, \alpha_2}
\, \left(  \pi^{,\beta_1}
/\sqrt{X}
\right)_{, \beta_2}
\, \left(  \pi^{,\delta_1}/\sqrt{X}\right)_{, \delta_2}
\,\,.
\eea
 The analogy with Galileon Lagrangians as expressed with the Levi-Civita symbols (see e.g. \cite{Fairlie:1991qe,Fairlie:1992nb} for early papers on these topics) is apparent. The traces of the
  tensors $Y^{{\bf (n)}}_{\,\mu \,\nu}$ are   total derivatives by themselves -- they lead to non-trivial
 theories only when weighted by the overall $\sqrt{X}$-coefficient in eq. \eqref{LagN}.
 
 Thanks to the antisymmetric properties  the  Levi-Civita symbols,  the scalar equations of motion are at most second order, as it happens for generalized Galileons (see e.g. \cite{Deffayet:2013lga,deRham:2012az} for  reviews).  Once applying
 the SL infinitesimal  transformation \eqref{dessym}, one finds the relations:
 
 \bea
 \delta_{\rm{sl}} \,\sqrt{X}&=&
   \omega^\rho\,\partial_\rho \left( \pi \,\sqrt{X}\right)
 \,,
 \\
 \delta_{\rm{sl}} \left(  \frac{\pi_{,\mu}}{ \sqrt{X}}\right)_{,\nu}&=&\,
 \omega^\rho\,\partial_\rho \left(\pi\,\frac{\pi_{,\mu}}{\sqrt{X}} \right)_{,\nu}
 \,.
 \eea
 Making use of these results, of the chain rule, and  the   structure \eqref{LagN} of the Lagrangians, it is
   straightforward  to realise that they transform  under
 the  infinitesimal SL transformation \eqref{dessym} as
 \be
 \delta_{\rm{sl}}\, {\cal L}_{n}\,=\,\omega^\rho\,\partial_\rho\, \left(
 \pi\,{\cal L}_{n}
 \right)\,,
 \ee
 for $n=1,\dots4$. 
 This
  is a total derivative,   showing that the SL transformation  \eqref{dessym} is a symmetry for these theories. After the work \cite{Chagoya:2016inc}, these
 symmetries have been more recently reconsidered in related contexts in \cite{Pajer:2018egx,Grall:2019qof}, and recognized to be specific of
 the cuscuton  Lagrangian   $\sqrt{X}$  as  eq. \eqref{L1fl}. Here we emphasize  that they are common to 
 all the  Lagrangians  above \eqref{L2fl}-\eqref{L4fl}. 
 
 \subsubsection*{Spontaneous breaking of Lorentz symmetry, and remarks on stability  under \\radiative corrections}

Besides their symmetries, another important feature of  
 Lagrangians  \eqref{L1fl}-\eqref{L4fl}  is 
  their specific structure: they are non-polynomial functions of $X$ (containing factors as 
 $\sqrt{X}$, $1/X$), implying that  
 the corresponding equations of motion {\it  do not allow} for  solutions with $X=0$. 
 A consequence is that these systems spontaneously break Lorentz symmetry: although  
   their Lagrangians are Lorentz invariant, the theories require  vacua with  a non-vanishing
   gradient for the scalar, with 
   $\bar \pi_{,\mu}\neq 0$ (a well-defined square root $\sqrt{X}$  further requires $\bar \pi_{,\mu}\,\bar \pi^{,\mu}\,<\,0$).
   Scalar backgrounds that depend only on time, $\bar \pi\,=\,\bar \pi(t)$,  are special:  
    making this choice of homogeneous background, the Lagrangians  \eqref{L3fl} and \eqref{L4fl}
   vanish identically, while \eqref{L1fl} and \eqref{L2fl} become linear  
       functions of the time-derivatives $\dot{ \bar \pi}$. Any linear
   combination of them with constant coefficients  then reduces to a total derivative, 
    ensuring that the homogeneous scalar configuration  $\bar \pi(t)$   automatically satisfies   the (trivial) equations
  of motion in Minkowski space and represents a consistent vacuum for the theory.  For this reason, it is natural
  to concentrate on the choice $\bar \pi(t)$.  This property of the Lagrangians of 
  becoming linear in the scalar first derivative, once one focusses on homogeneous time-dependent
  backgrounds, remains true also when the scalar system is coupled with other fields: we shall
  study some its cosmological consequences in section  \ref{cc-sect}.

\medskip

Around homogeneous scalar backgrounds $\bar \pi(t)$ -- as we proved in section \ref{sec-flat-sym} -- no linearized scalar fluctuations  propagate, making these scalar theories interesting. What should we expect about the stability 
of the structure 
of these theories under quantum effects? 
  In the absence of  dynamical  fluctuations around homogeneous backgrounds, we might  think that the structure 
of the Lagrangians  is protected under  quantum corrections associated with scalar modes --
   simply because no scalars propagate.  When considering
couplings with additional fields, one can induce symmetry breaking effects that render dynamical 
the scalar,
   making  arguments of stability of the structure of the theory   more subtle,
  also due to the absence of Lorentz invariance.
    Symmetry principles might help, as for the 
    non-renormalization theorems that characterize Galileons \cite{Luty:2003vm,Nicolis:2004qq}.
  We defer this problem to a future work, although we  make some additional remarks about this point  in discussing cosmological models based on these theories in section \ref{cc-sect}. 

\medskip

  In what comes next,  we show
 that these scalar systems can arise in a variety of contexts, once continuous or global symmetries
 get broken in certain ways. Scalar modes -- coupled with extra fields -- can then arise as Goldstone
 bosons of broken symmetries. We start in section \ref{sec-ab} to discuss vector theories in Minkowski space with
 broken $U(1)$ symmetry, while in section \ref{sec-gra} we consider scalar-tensor theories of gravity.

 \section{Scalarless interactions from broken   gauge symmetries}\label{sec-ab}

In this section, we discuss how 
the scalarless SL symmetry of section \ref{sec-flat-sym} can  characterize the physics of scalar excitations
 associated with broken gauge symmetries for vector degrees of freedom in flat space, again in the absence of gravity.  This opens the possibility to study  covariant
 theories that break  explicitly gauge symmetries, without propagating scalar modes around
 appropriate backgrounds. For definiteness, we focus on the simplest case of a broken Abelian $U(1)$ gauge symmetry. 

\smallskip

We consider the vector Lagrangian around flat space:
\be \label{vecLt}
{\cal L}_{V}\,=\,-\frac14\,F_{\mu\nu}\,F^{\mu\nu}\,-V(A_\mu,\,\partial_\rho A_\nu)\,,
\ee
with $F_{\mu\nu}\,=\,\partial_\mu A_\nu-\partial_\nu A_\mu$ is the field strength of a vector field $A_\mu$, while $V(A_\mu,\,\partial_\rho A_\nu)$ is a vector potential. 
 The vector potential generically breaks the Abelian
gauge invariance
\be\label{U1gauge}
A_\mu \,
\to \,A_\mu +\partial_\mu \xi\,,
\ee
that is instead obeyed  by the vector field strength $F_{\mu\nu}$: this fact normally implies the propagation
of a  longitudinal scalar excitation, due to  the gauge symmetry breaking potential $V$. In some cases, 
in an appropriate limit the Goldstone boson Lagrangian enjoys Galilean symmetries: this
fact was first explored in \cite{Tasinato:2014eka,Heisenberg:2014rta,Tasinato:2014mia}. 
   Here we instead study possible
realisations of the SL symmetry \eqref{dessym}. For definiteness we  focus on 
the vector Lagrangian \eqref{vecLt} with a potential
\be
V\,=\,\beta\,m^3 \,\sqrt{-\eta^{\mu\nu}\,A_\mu \,A_\nu}\,,
\label{vecpot}
\ee
where $\beta$ is a dimensionless parameter, and $m$ a mass
  scale  providing the correct
dimensionality.  
  This potential explicitly breaks the $U(1)$ gauge symmetry \eqref{U1gauge}; moreover,  it is
  characterised by a square-root structure  similar to the  scalar
  Lagrangians investigated in section \ref{sec-flat-sym}. 
 
 \smallskip
 
 In fact, our aim here is to exploit this similarity: 
  given what we learned
  in the previous section,  can we find new symmetries
  associated with the vector Lagrangian \eqref{vecLt}, which can prevent the propagation
  of linearized scalar modes? To answer this question more transparently, it is  convenient to work with an
  equivalent system, making use of the  Stuckelberg trick.
   We can re-instate  the original gauge  symmetry   \eqref{U1gauge} introducing a scalar field $\pi$
    in the combination
    \be \label{stucom}
    m\,A_\mu\,\to\,m\,A_\mu+\partial_\mu \pi\,, 
   \ee
    and writing the  
  scalar-vector potential
   \be\label{lagveccus1}
V\,=\,\beta\,m^2\,\sqrt{-\eta^{\mu\nu}\,\left(m\,A_\mu+ \partial_\mu \pi \right) \left(m\,A_\nu+\partial_\nu \pi \right)}
\,.
\ee
 The system associated with the potential \eqref{lagveccus1} is now invariant under the Abelian gauge symmetry
 \bea
 A_\mu&\to&A_\mu+\partial_\mu \xi \label{gaugesc12a}\,,
 \\
 \pi&\to&\pi-m \,\xi \label{gaugesc12}\,,
 \eea
 for an arbitrary scalar function $\xi$. 
 The scalar-vector potential  \eqref{lagveccus1} is  distinct from the original one, eq \eqref{vecpot}. On
 the other hand, using the gauge symmetry \eqref{gaugesc12}  to fix the gauge
    $\pi=0$ one recovers the original Lagrangian \eqref{vecLt}, that is 
 therefore physically equivalent to \eqref{lagveccus1}.   
 
 \smallskip
 
 Here  we make use of the symmetries eq \eqref{gaugesc12a}, \eqref{gaugesc12} and conveniently adopt the transverse
 Lorenz gauge $\partial^\mu A_\mu\,=\,0$ for describing the vector field dynamics.  
At this stage,  the structure of the potential \eqref{lagveccus1}
 would seem to imply
  that three degrees of freedom propagate in this system: two degrees of freedom in a gauge vector $A_\mu$, one
 degree of freedom in the scalar $\pi$.  
 On the other hand, considering eq
 \eqref{lagveccus1} and
  taking the decoupling limit $m\to0$,
  $\beta\to\infty$ such that $\beta\, m^2$ is kept finite, we get the scalar Lagrangian
  \be
  {\cal L}_{\rm dec}\,=\,\beta\,m^2\,\sqrt{-\eta^{\mu\nu}\,\partial_\mu \pi\,\partial_\nu \pi }\,,
  \ee
  which describes self-interactions of the Goldstone boson associated with the broken symmetry. 
  Having taken $m\to0$
this  scalar Lagrangian  looses the Abelian gauge symmetry \eqref{gaugesc12}; on the other hand,  it enjoys the additional SL symmetry introduced in section \ref{symflat}, which
prevents the propagation of scalar modes around time-dependent backgrounds $\bar \pi (t)$ that solve the equations of motion and spontaneously break Lorentz symmetry.

\medskip

 Such decoupling limit argument suggests
that the initial scalar-vector  potential \eqref{lagveccus1} is invariant under a more general scalar-vector global symmetry, even
outside a decoupling limit. In fact, a straightforward
computation along the arguments of section \ref{symflat} shows that the transformations (with $\omega^\rho $ an infinitesimal constant vector)
\bea
\label{newvs1}
A_\mu &\to&A_\mu+\omega^\rho\,\partial_\rho \left( \pi \, A_\mu\right) \,,
\\
 \pi&\to& \pi +\frac12\,\omega^\rho \partial_\rho\,\pi^2\,,
\eea
  leave the full potential \eqref{lagveccus1} invariant up to total derivatives.  This global symmetry is independent from $m$, and holds {also outside}  the decoupling limit discussed above.
   
 \smallskip
  
  While we focussed so far on the specific square root `cuscuton' potential \eqref{vecpot}, these very same 
   considerations apply also to  the vector versions of all the covariant scalar Lagrangians 
    \eqref{L2fl}-\eqref{L4fl},
    which shall contain
   derivatives of the vector fields. 
     It is sufficient
  to substitute the scalar first derivatives with the vector $A_\mu$
   in those Lagrangians to determine covariant  vector potentials. 
    Since the resulting vector Lagrangians contain non-polynomial functions as square roots and inverses of $A_\mu A^\mu$,
  consistent vacua around which fluctuations can be studied require a non-vanishing {\it vev} for $A_\mu$,
  or the more convenient  St\"uckelberg combination \eqref{stucom}.  The actual structure of the time-dependent
  {\it vev} will depend on the details of the complete theory, and possibly on the couplings of the vector with
  charged matter: a complete discussion goes beyond the scope of this section. (We will study
  in more detail this topic in the next section, when coupling scalars with gravity.) On the other hand, we notice that
 the simplest   option is to choose the interactions so to select a time-dependent background for the St\"uckelberg field 
  $\bar \pi(t)$.
    Linearized excitations around such background are parameterised by hat quantities $A_\mu\,=\,\hat A_\mu$, $\pi(t\,\vec x)\,=\,\bar \pi(t)+\hat \pi(t,\,\vec x)$. 
  The scalar fluctuations $\hat \pi$ enjoy the residual symmetry \eqref{dessym2} which ensures that no scalar excitation
  propagate around this vacuum~\footnote{Notice that the vector kinetic term $-F_{\mu\nu}^2/4$
  in eq. \eqref{vecLt} does not satisfy symmetry \eqref{newvs1}. On the other hand, this term preserves by itself the Abelian gauge symmetry which ensures that it does not excite scalar modes.
  }.

   \medskip
  
  In fact, the inevitable spontaneous breaking of Lorentz invariance of the initially covariant systems is
  the key to prevent the propagation of linearized scalar excitations around flat space, by exploiting the scalarless  symmetry \eqref{dessym}. 
  Notice that  these results are different from the partially massless  vector theories of \cite{Deser:1983mm}, 
  where
   scalar excitations are forbidden by a new gauge symmetry in de Sitter space that arises when the vector mass
   is related with the de Sitter scale. On the other hand, it would be interesting to understand whether
   our arguments and   techniques can be applied to theories of massive gravity, leading to examples of partially massless
   massive gravity theories in systems that spontaneously break Lorentz symmetry~\footnote{See e.g. \cite{Tasinato:2012mf} for
   an example of  enhanced symmetry in a spontaneously broken Lorenz violating setting, for  a Fierz-Pauli theory of massive gravity.}.

  \section{Scalarless interactions from broken space-time symmetries}
  \label{sec-gra}

Besides broken gauge symmetries, 
another context where scalarless interactions can arise are theories  breaking space-time symmetries. 
In
fact,  previous investigations
  \cite{Chagoya:2016inc}
  based on 
  \cite{deRham:2010eu} 
studied the possibility of obtaining  these theories from broken global symmetries in 5-dimensional
brane-worlds.  In this section we analyse in more generality this topic, and the consequences of coupling
the scalar theories of section \ref{symflat} with gravity. Namely:
\begin{itemize}
\item In section \ref{sec-bw-emb} we review the arguments of  \cite{Chagoya:2016inc} and show that the
scalarless SL symmetry \eqref{dessym} arises as a non-linear realization of broken global symmetries in a 5-dimensional
brane-world context with  a fixed space-time geometry.
\item In section \ref{sec-gr} we turn on dynamical gravity, and consistently couple our scalar Lagrangians  \eqref{L2fl}-\eqref{L4fl}
 with tensor modes in  curved space.
 This process generally breaks the SL symmetry that only holds in flat space; nevertheless we show that the
 scalar-tensor system obeys a novel continuous symmetry, inherited from broken diffeomorphism invariance
 in the 5-dimensional brane-world model. Such new symmetry reduces to the SL symmetry in the flat-space limit.
\item In section \ref{sec-conse} we investigate the consequences of the new symmetry in the scalar-tensor system
for the number of propagating degrees of freedom. In particular, we demonstrate that no linearized scalar modes propagate around
space-times satisfying appropriate conditions, and then focus on the dynamics of fluctuations around spherically-symmetric Schwarzschild  black holes. We show that no additional degrees of freedom propagate besides the ones of GR. On the other hand, the dynamics of perturbations around black holes are  different with respect to GR, making these theories distinguishable from Einstein gravity.
\item  In section \ref{cc-sect} we focus on cosmology, and discuss the possible relevance of our theories 
for the infrared dynamics of gravity and for the dark energy problem. We show that the effective action
controlling homogeneous cosmological quantities acquire the form of a constrained system, and a constraint
associated with the scalar field leads to  expanding space-time geometries which are {\it insensitive} to the
value of the cosmological constant. Moreover, the system leads to self-accelerating configurations, with
the rate of acceleration proportional to the parameters entering the scalar-tensor interactions.
\end{itemize}

 \subsection{Embedding in flat 5-dimensional brane-world models:\\
 the scalar as Goldstone boson of broken global symmetries}
 \label{sec-bw-emb}

A compelling physical motivation for 
the symmetry \eqref{dessym}  and
 the Lagrangians 
of eqs  \eqref{L2fl}-\eqref{L4fl} 
 arises from a brane-world perspective. This brane-world point of view was first developed in
  \cite{Chagoya:2016inc}, building on the results of
  \cite{deRham:2010eu} who
   introduced
 the DBI-Galileons. Here
 we briefly review this viewpoint  \cite{deRham:2010eu,Chagoya:2016inc} , and show that the scalar symmetry \eqref{dessym}  
   arises as a non-linear realisation of a broken global space-time symmetry in extra-dimensions.

We consider a 4-dimensional brane whose world-volume metric, $\gamma_{\mu\nu}$, is  embedded in a 5-dimensional bulk  whose metric reads

 \be\label{5dmetr}
d s^2_{5d}\,=\,{\bf g}_{M N}\,d X^M d X^N\,=\,\kappa_0\,\gamma_{\mu\nu} \,d x^\mu d x^\nu+ d y^2\,,
\ee
where  $y$ is the spatial extra-dimension.
 We introduce the  
   constant  parameter  $\kappa_0$  as a constant `warp factor' that plays an important role for our arguments.
     We   choose a  foliation of the 5-dimensional space expressing
 the brane embedding as
\bea
X^\mu&=&x^\mu\,,
\\
X^5&=&\pi(x)\,.
\eea
The scalar $\pi(x)$ geometrically parameterise fluctuations of the brane position in the fifth dimension. 
We
then
  introduce four 5-vectors as
\bea
e^{\,\,M}_{\mu}&
=&\frac{\partial\,X^M}{\partial\,x^\mu}\,,
\nonumber 
\\
&=&\left( ..\dots,\delta_\mu^{\,\,M},\dots,\,\partial_\mu \pi \right)
\label{four5v}\,.
\eea
(For example, $e^{\,\,M}_{0}\,=\,(1,0,0,0,\partial_0 \pi)$.)
 These four 5-vectors allow us to build the 
 induced metric on the brane (or first fundamental form)
as
\bea
g_{\mu\nu}(x)&\equiv&e^{\,\,M}_{\mu}\,e^{\,\,N}_{\nu}
{\bf g}_{M N}
\,
=\, \kappa_0\,\gamma_{\mu\nu}(x)+\partial_\mu \pi(x)\,\partial_\nu \pi(x)\,.
\label{def_ind_metric}
\eea
Introducing a unit normal vector to the brane $n^M$, orthogonal to the vectors \eqref{four5v},
 one  builds the second fundamental form -- usually called extrinsic curvature -- as the tensor
\be
{\cal K}_{\mu\nu}\,=\,e^{\,\,M}_\mu\,e^{\,\,N}_\nu\,\nabla_M\,n_N\,.
\ee
These are the tools   needed to build  the scalar  theories we are investigating. Indeed,
 as proposed in \cite{deRham:2010eu},  one can  consider   those
  Lagrangian densities  on the brane world-volume  
   that lead to second
  order equations of motion for the fields involved.
    These are  the  Lovelock invariants \cite{Lovelock:1971yv} with the addition of Gibbons-Hawking-York
  terms on the brane \cite{York:1972sj,Gibbons:1976ue}, leading to the actions
  \bea \label{loveinv}
  {\cal S}_1&=&\frac{1}{\kappa_0^3}\,\int d^4 x\,
    \sqrt{-g}\,,
    \\
      {\cal S}_2&=&\frac{1}{\kappa_0^2}\,\int d^4 x\,\sqrt{-g} \,{\cal K}
      \,,
\\
      {\cal S}_3&=&\frac{1}{\kappa_0}\,\int d^4 x\,\sqrt{-g} \,R
\,,
\\
      {\cal S}_4&=&\int d^4 x\, \sqrt{-g} \,{\cal K}_{\rm GB}
\,,
 \eea
 
 \smallskip
 
 \noindent
 where ${\cal K}$
  is the trace of the extrinsic curvature, $R$ the Ricci scalar computed with the induced
  metric $g_{\mu\nu}$,  while ${\cal K}_{GB}$, the `boundary-term' for the 5-dimensional Gauss-Bonnet combination \cite{Davis:2002gn}, is given by ${\cal K}_{GB}\,=\,{\cal K} \,{\cal K}_{\mu\nu}^2
  -2\,{\cal K}_{\mu\nu}^3/3-{\cal K}^3/3-2\,G^{\mu\nu} \,{\cal K}_{\mu\nu}$, with
  $G_{\mu\nu}$ the Einstein tensor evaluated with the metric $g_{\mu\nu}$. 
(The reason of the powers of $\kappa_0$ as overall coefficients will be clear soon.)
  
  \smallskip
  
  When expressing the actions of   \eqref{loveinv} in terms
  of the quantities $\gamma_{\mu\nu}$ and $\pi$ that appear in the expression
  for the induced metric \eqref{def_ind_metric}, one obtains covariant scalar-tensor theories whose
    equations of motion are automatically at most  second order.   
    The scalar field $\pi$, in this perspective,
plays the role of {Goldstone boson} of broken global symmetries, that
are non-linearly realised in the scalar theory under examination. 

 We now fix  $\gamma_{\mu\nu}(x)\,=\,\eta_{\mu\nu}$: the resulting scalar theories
 one obtains from Lagrangians \eqref{loveinv} are  the DBI-Galileons of
\cite{deRham:2010eu}. 
While in this section we
focus on broken global symmetries, in   next sections we shall examine how the scalars 
can non-linearly realise broken
 continuous symmetries in curved space-time. 
In this global case, the broken translation symmetry in the bulk is  non-linearly 
realised in terms of shift-symmetry acting on the scalar field. Broken 5-dimensional rotations are non-linearly realized in terms
of the infinitesimal symmetry transformation
\be \label{geninftr1}
\pi\,\to\,\pi\,+\,\kappa_0\,\omega_\mu x^\mu\,+\,\pi\,\omega_\mu\,\partial^\mu \pi \,,
\ee
for an arbitrary infinitesimal vector $\omega_\mu$. In the limit $\kappa_0\to0$, the 
symmetry \eqref{geninftr1} boils down to the infinitesimal SL symmetry introduced in eq
\eqref{dessym}. 

 In fact,
  the limit $\kappa_0\to0$  can be taken consistently (when including the appropriate powers of $\kappa_0$ as overall coefficients, as in eq \eqref{loveinv}) and 
  leads 
 to the  system 
 \eqref{L1fl}-\eqref{L4fl}. As shown in \cite{Chagoya:2016inc}, these  correspond to a particular case 
 of DBI-Galileon actions,  equipped with the  symmetry  \eqref{dessym} (or equivalently
 the $\kappa_0\to0$ limit of \eqref{geninftr1}). We refer the reader to \cite{Chagoya:2016inc} for
 more details on  $\kappa_0\to0$ limit of the geometrical brane-world construction with a flat 
 brane world-volume, and the associate global symmetries.  In the next sections, we turn on 
 gravity, to study how the scalar $\pi$ can be related with  broken continuous space-time symmetries.

\subsection{Coupling the scalar theory with gravity: a new symmetry arises}\label{sec-gr}

As first noticed in \cite{deRham:2010eu},
a consistent coupling with gravity for the system discussed in the  section \ref{sec-bw-emb} is straightforward: since actions \eqref{loveinv} are built in terms of Lovelock invariants and Gibbons-Hawking--like terms,
they are ensured to lead to second order equations of motion when expressed in terms of the constituents $\gamma_{\mu\nu}, \pi$ of the brane induced metric. 
 In fact,  taking the $\kappa_0\to0$ limit of actions  \eqref{loveinv}, 
one obtains special cases
 of Horndeski Lagrangian densities, which control possible  interactions between the metric $\gamma_{\mu\nu}$ and
 the scalar $\pi$:

\bea
{\cal L}_1
 &=&
 \sqrt{X}\label{Lag1t}
\,,
\\
{\cal L}_2
 &=&
 \frac{  \pi_{;\mu}\,\Pi^{\mu\nu}  \, \pi_{; \nu}}{{X}}
\label{Lag2t}
\,,
\\
{\cal L}_3
 &=&
R+\frac{1}{{X}}\left([\Pi^2]-
[\Pi]^2
\right)
\label{Lag3t}
\,,
\\
{\cal L}_{ 4}
&=&
\frac{1}{X}\,
\left([\Pi]^3+2\, [\Pi^3] -3 \,[\Pi^2] [\Pi]-{12}\,
\Pi_{\mu}^{\,\,\rho}\,\pi_{;\rho}
\,G^{\mu\nu}\,\pi_{;\nu}\right)
 \label{Lag4t}
\,,
\eea

\noindent
where $\pi_{;\mu}\,\equiv\,\nabla_\mu \pi$, $\Pi_{\mu\nu}\,\equiv\,\nabla_\mu \nabla_\nu \pi$, 
$X\,\equiv\,-\pi^{;\mu} \,\pi_{;\mu} $, and indexes are raised/lowered
with the dynamical metric $\gamma_{\mu\nu}$, which is also used to take covariant derivatives. 
The Ricci scalar $R$ and Einstein tensor $G_{\mu\nu}$ appearing in the previous
expressions are computed in terms of the induced 4-dimensional metric $\gamma_{\mu\nu}$.

As anticipated above,  
these theories lead to second order equations of motion, and belong
to the class of Horndeski actions.
  Besides Horndeski systems, also special cases of beyond Horndeski or DHOST theories \cite{Langlois:2015cwa,Langlois:2015skt,Crisostomi:2016czh,deRham:2016wji} can
  be obtained in terms of brane-induced scalar quantities, as pointed out in \cite{Chagoya:2018yna}. In particular, systems
  built in terms of  the trace of powers
  of the extrinsic curvature tr$\left[{\cal K}_{\mu \nu}^{n}\right]$, $n\ge2$ are known to lead to consistent covariant
  theories which avoid Ostrogradsky instabilities \cite{Crisostomi:2016czh,Crisostomi:2016tcp}.  Before starting
  to analyse the properties of these systems, we can make some general comments:
    \begin{itemize}
  \item  The Lagrangians \eqref{Lag1t}-\eqref{Lag4t} contain non-polynomial functions of $X$ as $\sqrt{X}$, $1/X$,
   hence they spontaneously break Lorentz invariance, since
    consistent solutions of the equations of motion require a non-vanishing gradient for the scalar $\partial_\mu \bar \pi \neq 0$. In what follows we will be interested to explore the consequences of this fact.
      \item 
 The scalarless symmetry \eqref{dessym} is expected to break around curved space-times, since 
  a covariantly constant vector $\omega^\mu$ generally does not exist outside flat space. Nevertheless,
  we shall learn that
   the
 scalar-tensor system obeys a novel continuous symmetry, inherited from broken diffeomorphism invariance
 in the 5-dimensional brane-world model.

  \end{itemize}

\subsubsection*{A  local symmetry inherited from higher-dimensions}

The covariant scalar-tensor theories
discussed above are expected to break symmetry \eqref{dessym}  around generic
curved configurations; on the other hand, we now discuss an additional symmetry
for these systems, inherited from 5-dimensional bulk continuous symmetries.
This symmetry was already noticed in \cite{Chagoya:2018yna}, without however
 realizing its higher dimensional origin.

We focus on the same 5-dimensional brane-world perspective introduced in section \ref{sec-bw-emb}. 
 We consider 
   the continuous symmetry associated with
reparameterisation 
invariance of the bulk space-time,   broken  by the presence of the brane. The bulk coordinate
transformations read:

\be\label{coort1}
X^M\,\to\,X^M-\chi^M(X^N)\,,
\ee
with $\chi^M$ denoting a 5-dimensional vector with 5 independent components. 
The 5-dimensional metric transforms as a tensor under this symmetry, which leads 
to the infinitesimal transformation
\bea
{\bf g}_{MN}( X)&\to&\frac{\partial \tilde X^P}{ \partial  X^M}\frac{\partial \tilde X^Q}{ \partial  X^N}\,{ \bf g}_{PQ}(\tilde X)
\nonumber
\\
&=&
 {\bf g}_{MN}+\chi^L\,\partial_L\,{\bf g}_{MN}+\partial_M \chi^L\,{\bf g}_{LN}
+\partial_N \chi^L\,{\bf g}_{LM}\,.
\eea

 We get non-linearly realised induced symmetries in
 the brane world-volume,
  controlled by the components
of the 5-vector $\chi^A$.
 We concentrate on
   vectors  $\chi^M$ depending only on 4-dimensional coordinates
\be
 \chi^M\,=\,\left( \xi^\mu(x),\,\kappa_0 \psi(x) \right)\,,
\ee
 with $\psi(x)$ an arbitrary scalar field. This choice is made
 to preserve the standard 4-dimensional  diffeomorphism transformations. In fact, the first four components $\xi^\mu$ act on the induced metric
 $\gamma_{\mu \nu}$ and the scalar $\pi$ exactly as standard coordinate invariance of General Relativity, which we
 assume to
 hold in the covariant scalar-tensor system. The component $ \chi^5\,=\, \kappa_0 \,\psi$ provides us with a new symmetry
 transformation, that acts on the metric and the scalar as

\bea
g_{\mu\nu}(x)
 &\to&g_{\mu\nu}+\left( e_\mu^{\,\,\rho}\,e_\nu^{\,\,5}+ e_\nu^{\,\,\rho}\,e_\mu^{\,\,5}
\right)\,\partial_\rho\,\chi^5\,{\bf g}_{55}
\nonumber
\\
&=&\kappa_0\,\left( \gamma_{\mu\nu}+\partial_\mu \psi \partial_\nu \pi 
+\partial_\mu \pi \partial_\nu \psi 
\right)+\partial_\mu \pi\,\partial_\nu \pi
\,,\label{contsym1}
\eea
while
\bea
\pi(x)&\to& \pi(x)-\kappa_0\,\psi(x)\,.
\eea

 In the limit $\kappa_0\to0$ that, as
we learned  in section \ref{sec-bw-emb},
   is of most interest for us the transformation acts on the tensor field $\gamma_{\mu\nu}$ only, and reads
 \be\label{locnews}
 \gamma_{\mu\nu}\to \gamma_{\mu\nu}+\partial_\mu \psi \partial_\nu \pi 
+\partial_\mu \pi \partial_\nu \psi \,,
 \ee
for an arbitrary, infinitesimal  scalar function $\psi(x)$.  The transformation \eqref{locnews}
acting on the metric (and not on the scalar) is then a  continuous infinitesimal  symmetry
for  the covariant scalar-tensor actions 
\eqref{Lag1t}-\eqref{Lag4t}. 

As a concrete example, we consider the cuscuton system \eqref{Lag1t} minimally coupled with gravity,
which we rewrite here
\be
S_1\,=\,\int d^4 x\,\sqrt{-\gamma}\,\sqrt{X}\,.
\ee
 The argument of the integral   transforms under the symmetry as 
\bea
\sqrt{-\det{\gamma_{\mu\nu}}}\,\sqrt{X}&\to&\sqrt{-\det{\left(\gamma_{\mu\nu} +\partial_\mu \psi \partial_\nu \pi 
+\partial_\mu \pi \partial_\nu \psi \right)}}\,\sqrt{-\left(\gamma^{\mu\nu} -\partial^\mu \psi \partial^\nu \pi 
-\partial^\mu \pi \partial^\nu \psi\right)\,\partial_\mu \pi \partial_\nu \pi}\nonumber
\\
&=&\sqrt{-\det{\gamma_{\mu\nu}}}\,\left(1+ \partial \pi\cdot \partial \psi\right)\,
\sqrt{X}\,\left(1-\partial \pi\cdot \partial \psi\right)
\\
&=&\sqrt{-\det{\gamma_{\mu\nu}}}\,\sqrt{X}\,,
\eea
where we kept terms up to linear order in the small scalar quantity $\psi$, showing as stated that action
\eqref{Lag1t} is invariant under symmetry \eqref{locnews}.

\subsubsection*{Relating curved and flat space symmetries in the Minkowski limit}

We now argue that  symmetry \eqref{locnews} might be regarded as a  scalar-tensor, curved-space version of 
the infinitesimal scalar symmetry \eqref{dessym}  introduced around flat space. To do so, 
   we  
focus  on
   fluctuations around a  Minkowski geometry, and assume a   Lorentz-violating scalar background $\bar \pi$
 whose gradient has non-vanishing norm,
 say $\partial_{\nu} \bar \pi\,\partial^{\nu} \bar \pi\,<\,0$; for definiteness, we choose $\bar \pi\,=\,t$.
 We aim to then relate  \eqref{locnews} with  \eqref{dessym}.
  Symmetry \eqref{locnews}
acting on infinitesimal metric fluctuations $h_{\mu\nu}$ around flat space (that is,  $\gamma_{\mu\nu}\,=\,\eta_{\mu\nu}+h_{\mu\nu}$)
reads
\be\label{infsym1}
h_{\mu\nu}\,\to\, h_{\mu \nu}+\partial_{\mu} \left( \psi\, \partial_{\nu} \bar \pi\right)
+\partial_{\nu} \left( \psi\,  \partial_{\mu} \bar \pi\right)\,,
\ee
which is equivalent to an infinitesimal coordinate transformation
\be\label{infsym1a}
x^\mu\to x^\mu+ \psi \,\partial^\mu \bar \pi\,,
\ee
for an arbitrary infinitesimal scalar quantity $\psi$. We stress, on the other hand, that the specific  transformation \eqref{infsym1a} only acts on the tensor fluctuations $h_{\mu\nu}$, and not on the scalar field. 

We can  exploit the similarity of \eqref{infsym1} with diffeomorphisms   to `transfer' the symmetry transformation 
  from the tensor to the scalar sector: we apply a standard infinitesimal diffeomorphism transformation to the metric fluctuation
\be\label{infsym2}
h_{\mu\nu}\,\to\,h_{\mu\nu}+\partial_\mu \xi_\nu +\partial_\nu \xi_\mu \,,
\ee
and choose
$
\xi_\mu\,=\,- \psi \,\partial^\mu \bar \pi\,$. 
The combination of the two transformations \eqref{infsym1} and  \eqref{infsym2} compensates each other, leaving
the metric fluctuations unchanged; on the other hand, the diffeomorphism transformation  in   \eqref{infsym2} acts
on the scalar fluctuations as well, 
\be
\hat \pi \to \hat \pi+\xi^\mu\,\partial_\mu \bar \pi\,=\,\hat \pi- \psi \,
\left(
\partial^\mu \bar \pi \partial_\mu \bar \pi
\right)\,.
\ee
By
 choosing the free function $\psi$ as 
$\psi 
 \,=\,\left( \bar \pi \omega^i\,\partial_i\hat \pi\right)/\bar X$, this transformation becomes identical
to the flat-space residual symmetry  \eqref{dessym2}
acting on scalar fluctuations. In this sense, 
 the symmetry acting on the metric \eqref{locnews} is
 a scalar-tensor generalization of the flat-space scalar symmetry we started with in section \ref{sec-flat-sym}.

  \subsection{Consequences  of the symmetry for  the   propagating degrees of freedom}\label{sec-conse}

We 
 investigate the 
consequences of symmetry \eqref{locnews} for the number of propagating degrees of freedom
around curved space-times, extending the arguments of 
\cite{Chagoya:2018yna}.

Theories \eqref{Lag1t}-\eqref{Lag4t} lead to  scalar background configurations that solve
the equations of motion
spontaneously breaking Lorentz symmetry, via a space-like gradient for the background 
scalar $\bar \pi(x)$ such that $\partial_\mu\,\bar \pi\,\neq\,0$. 
 We study the dynamics of linearized fluctuations around background solutions. 
We split 
scalar and metric fluctuations as
\bea
\pi &=&\bar \pi+\hat \pi\,,
\label{flupi1}
\\
\gamma_{\mu\nu} &=&\bar \gamma_{\mu\nu}+\hat h_{\mu\nu} 
\,,
\label{flucur1}
\eea
where the hat quantities are infinitesimal. 
Symmetry \eqref{locnews} does not act on the scalar (hence it leaves invariant the scalar background $\bar \pi$) and acts
on metric fluctuations as
\be\label{symme1}
\hat h_{\mu\nu}\,\to\,\hat h_{\mu\nu}+\bar \nabla_\mu {\hat \psi}\,\bar \nabla_\nu \bar \pi
+
\bar \nabla_\nu {\hat \psi}\,\bar \nabla_\mu \bar \pi
\ee
with ${ \hat \psi}(t\,\vec x)$ an arbitrary infinitesimal scalar field. In certain circumstances,  symmetry \eqref{symme1} ensures
that scalar excitations do not propagate.  To see this fact explicitly, we work with the action
\be\label{totacs1}
S_{\rm{}}\,=\,\int d^4 x\,\sqrt{-\gamma}\,\left[ M_{\rm {Pl}}^2\,R+\sum_i\,c_i\,{\cal L}_i \right]\,,
\ee
where ${\cal L}_i$, $i=1,\dots4$ are the Lagrangian densities \eqref{Lag1t}-\eqref{Lag4t}, and $c_i$ constant coefficients. 
We add for completeness the Einstein-Hilbert term $M_{\rm {Pl}}^2\,R$ that of course by itself
does not introduce scalar excitations. 
We select
background configurations that are solutions to the corresponding equations of motion with
non-vanishing gradient $\bar \nabla_\mu \bar \pi$, and study the 
dynamics of small fluctuations defined as  in eqs \eqref{flupi1}, \eqref{flucur1}. The total action \eqref{totacs1} is invariant under
diffeomorphisms: 
\bea
\label{metrtran1}
\hat h_{\mu\nu}&\to& \hat h_{\mu\nu}+\bar \nabla_\mu \hat \xi_\nu+\bar \nabla_\nu \hat \xi_\mu
\,,
\\
\hat \pi&\to&\hat \pi+\hat \xi^\rho\,\bar \nabla_\rho\,\bar \pi
\,,
\eea
for an infinitesimal vector $\hat \xi^\mu$ (the covariant derivatives are taken with respect to the 
background metric $\bar \gamma_{\mu\nu}$). 
We can use this fact to `gauge-away' the scalar fluctuations by selecting the  gauge-fixing vector
\bea
\hat \xi_{\rm gf}^\mu&=&-\hat \pi\,
\left(
\frac{\bar \nabla^\mu \bar \pi}{\bar \nabla^\rho \bar \pi\,\bar \nabla_\rho \bar \pi}
\right)\,
\equiv\,-\hat \sigma \,\bar \nabla^\mu \bar \pi
\eea
where, for convenience, we collect pieces depending on the infinitesimal parts of the transformation into
the scalar combination $\hat \sigma$. 
This choice of diffeomorphism vector $\hat \xi^\mu_{\rm gf}$ breaks diffeomorphism invariance fixing a gauge: while this is not an  issue for the
Einstein-Hilbert  contribution to action \eqref{totacs1} -- which we know propagate only 2 tensor modes -- it can be a problem for    Lagrangians ${\cal L}_i$: such gauge choice `moves'
scalar excitations from the scalar   to the metric sector through the gauge transformation  \eqref{metrtran1},
which reads here
\bea
\label{metrtran2}
\hat h_{\mu\nu}&\to& \hat h_{\mu\nu}-\bar \nabla_\mu\left(\hat \sigma \,\bar \nabla_\nu \bar \pi\right)
-
\bar \nabla_\nu \left(\hat \sigma \,\bar \nabla_\mu \bar \pi\right)\,.
\eea

 On the other hand, recall that the scalar-tensor Lagrangians are invariant under symmetry \eqref{locnews}: in cases where the scalar gradient $\bar \nabla_\mu \bar \pi$ is a Killing vector for the geometry
\be\label{kilcon}
\bar \nabla_\mu\,\bar \nabla_\nu\,\bar \pi\,=\,0\,,
\ee
then -- when choosing $\hat\psi\,=\,\hat \sigma$ --  the 
 the symmetry transformation \eqref{symme1} exactly compensates 
diffeomorphism transformation \eqref{metrtran2}, leaving
the metric fluctuations unchanged. Hence, for background configurations
satisfying the Killing condition \eqref{kilcon} our symmetry arguments ensure that no scalar fluctuations propagate.

\smallskip

 It is important to emphasize that our arguments do not imply that {\it only configurations satisfying \eqref{kilcon} avoid the propagation of scalar modes}.
 In fact, there are studies based on Hamiltonian analysis that show that
 theories as ours do not propagate scalar modes, once the scalar profile satisfies a unitary gauge, with
 no hypothesis on the metric profile \cite{Iyonaga_2018,Gao_2020}. In fact, our arguments provide
   sufficient conditions based on symmetries, not  necessary ones: there can very well be situations where scalar modes do not propagate thanks to second-class constraints. Such second-class constraints might get `promoted' to first-class -- and then to local gauge symmetries -- around certain configurations, as we are finding here.

\smallskip

We now proceed making a concrete example of spherically-symmetric configurations
where the symmetry prevents the propagation
of scalar excitations, nevertheless leading to differences with respect to GR for what respect
propagating tensor modes.

\subsubsection*{Spherically symmetric configurations}

The specific theories we examine admit various branches of spherically symmetric configurations, which
have been studied in \cite{Chagoya:2018yna}. Solutions exist
whose metric is   identical to the ones of GR, and the scalar acquires a time-dependent
profile, as in the configurations first discussed in \cite{Babichev:2013cya,Kobayashi:2014eva}. Including
for definiteness only Lagrangian ${\cal L}_4$ in  \eqref{totacs1}, we consider the action
\bea\label{totac2}
S\,=\,{M_{\rm Pl}^2}\,\int d^4 x\,\sqrt{-\gamma}\,\left[
 R+ \lambda\,\sqrt{X}\,
 \,\left( R+\frac{1}{{X}}\left([\Pi^2]-
[\Pi]^2\right)\right)
\right]\,,
\eea
 corresponding to  a special case of quartic Horndeski, with $G_4\,=\,1+\lambda\,\sqrt{X}$; when 
 choosing the dimensionless constant parameter $\lambda=0$
we recover the standard Einstein-Hilbert action.  This theory admits a spherically
symmetric solution whose metric corresponds to a Schwarzschild black hole; for our purposes it is convenient
to express it in terms of Lemaitre coordinates, as proposed in the recent work \cite{Khoury:2020aya}
\bea
\bar \pi&=&\bar\Lambda \tau\,, \label{solss1}
\\
d s^2&=&-d \tau^2+\frac{r_s}{r} d \rho^2+r^2\,d \Omega_2^2\,,\label{solss2}
\eea
where $r_s\,=\,2 M$, $\bar \Lambda$ is an energy scale that from now on we set equal to one, and  the quantity $r$ is expressed in terms of time and radial coordinates $(\tau,\,\rho)$
as
\be
r\,=\,r_s^{1/3}\,\left[\frac32\,\left( \rho-\tau\right) \right]^{2/3}\,.
\ee
The Horndeski function $G_4$ reads for this configuration 
$\bar G_4\,=\,1-\lambda\,\gamma^{00}\,\left(\partial_0 \pi\right)^2 \,=\,
1+\lambda
$. (Recall that we take $\bar \Lambda=1$).
The constant parameter $\lambda$ controls deviations from GR. 
A simple change of coordinates allows one to express the previous configuration in standard Schwarzschild coordinates; on the other hand, for our purposes it is convenient to make use of  Lemaitre coordinates in which
the scalar profile only depends (linearly) on time, and whose gradient satisfies the Killing condition \eqref{kilcon}.

 Around
the scalar-tensor configuration \eqref{solss1}, \eqref{solss2}, then, no scalar excitations propagate for the arguments we developed above, but only the spin-2 tensor excitations of GR. On the
other hand, the dynamics of linear fluctuations change with respect to GR, making this scalar-tensor
theory distinguishable with respect to Einstein gravity. The study of scalar and metric
fluctuations around  black hole backgrounds with time-dependent scalar profile was started in 
 \cite{Kobayashi_2012,Kobayashi_2014}, and  stability issues in the scalar sector have been discussed  in
  \cite{Ogawa_2016,Takahashi_2017,Khoury:2020aya,Babichev_2018,de_Rham_2019}.
   In our case, instabilities are absent because scalar modes do not propagate. 
   We study the dynamics of  tensor modes using the methods of \cite{Khoury:2020aya}, and relegate the technical analysis to appendix \ref{app-BHpert}.

   As a representative example,  
  we
write here the linearized Lagrangian governing the single parity odd-excitation $\Psi$ for multipoles $\ell\ge2$ (see appendix \ref{app-BHpert} for details):

\bea\label{rewhe1}
{\cal L}_{\rm odd}\,=\,\frac{1}{2\left(1+\lambda\right)}\sqrt{\frac{r_s}{r}}\frac{r^2}{\ell (\ell+2)-2}
\left[ \dot \Psi^2-\frac{(1+\lambda) r}{r_s}\Psi'^2+\left(\frac{15\,r_s}{4 r^3}
+\frac{(1+\lambda)}{4 r^2}\left( 3-4 \ell (\ell+1)\right)\right)\Psi^2
\right]
\nonumber\,,\\
\eea
where dot and prime indicate respectively derivatives along $\tau$ and $\rho$. When $\lambda=0$, this
Lagrangian density leads to the Regge-Wheeler equation for parity-odd black hole fluctuations
in Lemaitre coordinates. When $\lambda\neq0$, instead, we notice that fluctuations propagate
with a non-unit sound speed
\be
c_\Psi\,=\,1+\lambda\,,
\ee
and the Regge-Wheeler potential is also affected in the contributions proportional to the multipole numbers $\ell$. 
  For avoiding gradient instabilities, we must require $1+\lambda\,>\,0$.

This explicitly  demonstrates
 that black hole fluctuations can  allow one to distinguish  GR from a gravity theory controlled by action  \eqref{totac2}, even
 if they admit  spherically symmetric solutions identical to Einstein gravity and do  propagate
 the same degrees of freedom. While we discussed here odd-parity modes,
 also even-parity mode dynamics is sensitive to the parameter $\lambda$, and in appendix  
 \ref{app-BHpert}  we explicitly
 demonstrate 
   that scalar kinetic terms relative to scalar modes vanish.  This does not indicate strong coupling problems -- instead
   is a consequence of the symmetry \eqref{symme1} discussed above.

\subsection{Constrained cosmology and the dark energy  problem}
\label{cc-sect}

In this section we point out some additional features of the scalar-tensor systems
under considerations, which make them interesting for characterizing 
the infrared properties of gravity. In the structure of the  scalar-tensor interactions  \eqref{Lag1t}-\eqref{Lag4t}  
 the scalar $\pi$ formally appears with a {\it single power} in each of these
Lagrangians (in the loose sense that a square root as $\sqrt{-(\partial \pi)^2}$ reduces
the power from two to one).
 This structure is essential for developing the symmetry arguments
 as in the previous sections, but has the extra consequence that the scalar
 field, in certain circumstances, can act as {\it Lagrange multiplier} for the
 scalar-tensor set-up. Namely, its equation of motion can provide a constraint
 for the gravity sector of the theory, independent
 from additional couplings of gravity with other fields. This feature occurs when focussing
  in homogeneous 
  time-dependent configurations. 
  In fact, we already remarked in section \ref{symflat} (in absence of gravity) that when the scalar depends
  on time only, the Lagrangians become linear in $\dot \pi$, and this quantity becomes a constraint. When gravity is turned on, this linear
  dependence leads to a constraint condition for cosmological systems.
  We  investigate this property in a specific example,  aimed
 to  explore its implications for the dark energy  problem.
  
 \smallskip
 
 We focus on the action 
 \bea
S&=&\int d^4 x\,\sqrt{-\gamma}\,\Big\{ M_{\rm Pl}^2\,R-6\,\Lambda\,
\nonumber
\\
&&\hskip2cm
+\frac{\lambda}{2}\,\left[6\, H_0^2\,\sqrt{X}+\sqrt{X}\,R+\frac{2}{X^{3/2}} \left( \partial_\mu \pi 
\,\Pi^{\mu\rho}\,\Pi_{\rho \sigma}\,\partial^\sigma \pi-
\Pi\,\partial_\mu \pi 
\,\Pi^{\mu\rho}\,\partial_\rho \pi
\right)
\right]
\Big\}
\nonumber
\\
\label{dhostac}
\eea
The first line is the Einstein-Hilbert action equipped with a cosmological constant $\Lambda$; the second line
contains the interactions we are interested in, weighted by an overall dimensionless constant $\lambda$. They
include the cuscuton term $\sqrt{X}$, with an extra factor $H_0^2$ with dimension of mass squared,
 and a specific non-minimal coupling with gravity in the DHOST class, selected so to ensure
 that tensors propagate with the speed of light around curved background configurations. (Notice
 that these specific interactions do not respect the symmetry discussed in section \ref{sec-gr}, but this is
 not relevant for the arguments of this section: here we only make use of the structure of the actions discussed
 in section \ref{sec-gr}.) These interactions have the 
 property that the scalar appears linearly in the action. We specialise on time-dependent,
 homogeneous background with a FRW Ansatz for the metric
 \bea
 \pi&=& \bar \pi(t)\,,
 \\
 d s^2&=&-N^2(t)\,d t^2+a^2(t)\,d \vec x^2\,.
 \eea
We can  derive the following effective action for the time-dependent quantities $ \pi(t)$, $N(t)$, and $a(t)$, once the previous Ansatz is plugged in action \eqref{dhostac} 
\be
S\,=\,{\cal C} \,\int d t \,N(t)\,a^3(t) \left[ \frac{H^2(t)}{N^2(t)}+\frac{\Lambda}{M_{\rm Pl}^2}-
\frac{\lambda}{M_{\rm Pl}^2}\,
\frac{\dot{ \bar \pi}(t)}{2\,N(t)} \left( H_0^2- \frac{H^2(t)}{N^2(t)}\right)
\right]\,.
\ee
The overall constant ${\cal C}$ includes the 3-dimensional volume factor, and we denote $H\,=\,\dot a/a$. 
Both the lapse $N(t)$ and the scalar derivative $\dot \pi(t)$ act
as Lagrange multipliers for the system. The two independent field equations for the scale factor and $\pi$
read 
\bea
\label{Nco1}
H^2(t)\,\left(1+\lambda\,\frac{\dot{ \bar \pi}}{M_{\rm Pl}^2}  \right)&=&\frac{\Lambda}{M_{\rm Pl}^2}\,,
\\
H^2(t)+\frac23\,\dot H(t)&=&{H_0^2}
\label{scalco1}\,.
\eea
The second equation is a constraint associated with the scalar Lagrange multiplier: 
 it controls the Hubble parameter {\it independently} from the cosmological constant $\Lambda$. 
The solution for this system of equations
is
\bea
\label{Htsol1}
H(t)&=&H_0\,\tanh{\left[\frac{3\,H_0}{2} \left(t+t_0\right) \right]}\,\,\to\,\,H_0\,,
\\
\dot {\bar \pi}(t)&=& \frac{{-H_0^2 M_{\rm Pl}^2+\Lambda\,\coth^2{\left[\frac{3\,H_0}{2} \left(t+t_0\right) \right]} }}{{\lambda} \,H_0^2\ }\,\,\to\,\,\frac{{\Lambda\,
 -H_0^2 M_{\rm Pl}^2}}{{\lambda} \,H_0^2\ }\,,
\label{scalco2}
\eea
with $t_0$ an integration constant, and the arrows indicate
the late time asymptotics. At late times the Hubble parameter converges to the constant
value $H_0$ independent from $\Lambda$, as expected from the constraint
condition \eqref{scalco1}. Being a constrained field, the scalar profile accommodates
 as in expression \eqref{scalco2} so to solve eq \eqref{Nco1}.
 Notice that in our discussion we did not include additional contributions in the the form
 of matter or radiation; here we are only interested on cosmic acceleration, and we leave 
 a study of realistic cosmology to a separate work. We nevertheless point out  that  the scalar
 can couple linearly to matter fields, preserving the property of the constraint equation above.
  
\medskip

At what extent this mechanism is of use for addressing the dark energy problem? The ideas
 discussed above rely on two features, whose study  goes beyond the scope of this
work:
\begin{enumerate}
\item  As apparent
from eq \eqref{Htsol1}, the scale of late-time acceleration is controlled by the parameter $H_0$
entering as overall factor in the cuscuton contribution to action \eqref{dhostac}. If it were for this term only,
we would expect that the scalar is not dynamical thanks to the symmetries discussed in the previous
sections.
   The scalar, on the
other hand, couples to itself  through  DHOST contributions in 
\eqref{dhostac}, and possibly  to matter fields in a realistic cosmological setting, making it dynamical
 in an expansing space-time. Propagating  fields -- the
scalar, graviton, or matter fields -- can
induce quantum 
   loop corrections  that renormalize the scale $H_0$ in the cuscuton action. It would be interesting
to understand whether symmetries as the ones discussed in section \ref{symflat} can lead to non-renormalization
theorems similar to Galileons \cite{Luty:2003vm,Nicolis:2004qq} which can protect the overall coefficients in the Lagrangian, even when they are softly broken in a cosmological set-up \cite{Pirtskhalava:2015nla}.

\item 
Besides a mechanism of self-acceleration, a distinctive feature of the mechanism
above is that the 
  bare cosmological constant $\Lambda$ does not contribute to the rate of expansion,
   thanks to the constraint condition \eqref{scalco2}.
  This result seems relevant for addressing the cosmological constant problem (see e.g. \cite{Weinberg:1988cp,Padilla:2015aaa} for
  reviews). Whether it can lead to a realistic 
  solution to the problem without fine-tuning is an interesting question: we notice here 
  that this idea resembles the approach of 
   unimodular gravity to solve the cosmological constant problem, where a constraint
   condition compensates a cosmological constant contribution to the field equations.
    Unimodular gravity does {\it not solve} the cosmological constant problem since 
    operators that impose the constraint conditions (and control the rate of expansion) receive  
    large quantum contributions equivalent to the cosmological constant problem, as shown in \cite{Padilla:2014yea}. 
    In our case, the analog question arises with respect of the stability of the parameter $H_0$ under loop corrections,
    as mentioned above.
\end{enumerate}

\smallskip
\noindent
    We leave  futhere investigations on these points to future work.

\section{Outlook}\label{sec-out}

In this work we developed  the analysis of \cite{Chagoya:2016inc,Chagoya:2018yna}, discussing
how symmetry principles can be used for determining scalarless scalar interactions, which prevents
 the propagation of 
 scalar fluctuations around certain backgrounds solving the equations of motion. 
 We    understood how  scalarless interactions can arise when breaking gauge or space-time symmetries
   in vector or tensor theories, with the scalar playing the role of Goldstone bosons of such symmetries.
  We shown that such symmetries are generally associated with theories that spontaneously
break Lorentz invariance by a non-vanishing scalar field gradient. 
   In the scalar-tensor case, we provided sufficient conditions  for ensuring that scalar perturbations do not propagate
   around appropriate background configurations equipped by Killing vectors. 
  We then studied the consequences of our results for spherically symmetric configurations and for cosmology. We proved that scalar excitations do not propagate around black hole solutions; the dynamics of spin-2 modes, nevertheless, is different with respect to Einstein gravity, making these theories distinguishable from General Relativity. In a cosmological setting, we noticed  that these theories lead to constrained cosmological settings,
   with self-accelerating cosmological solutions in which the rate of cosmological
   expansion is independent from the value of a bare cosmological constant. 
   
   \smallskip
   We leave many interesting questions to future work:
   \begin{itemize}
   \item Symmetry principles can     limit the number of allowed interactions for the set-up
 under examination, and protect them under classical
   and quantum corrections. It would be interesting to understand how symmetries, even when 
   broken in a space-time set-up as in section \ref{cc-sect}, can protect the size and structure
   of  operators useful for addressing the dark energy problem. 
   \item 
   Besides solutions  with Killing vectors, see section \ref{sec-conse}, it would
   be interesting to find other contexts where the space-time symmetry
   of section \ref{sec-gr} can be used to prevent the propagation of scalar fluctuations. A
   possibility is to use 
    the high/low frequency splitting
   to study gravitational wave propagation in scalar-tensor theories as in \cite{Garoffolo:2019mna}, and study whether
   symmetries acting on high-frequency modes forbid the propagation of scalar fluctuations.
   \item  While in section \ref{sec-ab} we studied  broken Abelian vector theories that do 
   not propagate scalar modes, it would be interesting to investigate whether partially massless massive
   gravity theory exists, in systems that spontaneously break Lorentz symmetry.
   \end{itemize}
   
   \section*{Acknowledgments}

It is a pleasure to thank Ivonne Zavala for her careful reading of the manuscript. GT is partially funded by STFC grant ST/P00055X/1.

\begin{appendix}

\section{Scalarless perturbations around black-hole space-times}\label{app-BHpert}
In this appendix we further develop the arguments of section \ref{sec-conse}.
 Using methods and results of \cite{Khoury:2020aya}, 
 we study fluctuations around the spherically symmetric black hole
configurations,  that 
 are solutions to the field equations with spherically symmetric Ansatz associated with the action  (we set here $M_{\rm Pl}\,=\,1$)
\bea
S\,=\,\int d^4 x\,\sqrt{-\gamma}\,\left[
 R+ \lambda\,\sqrt{X}\,
 \,\left( R+\frac{1}{{X}}\left([\Pi^2]-
[\Pi]^2\right)\right)
\right]\,,
\label{actAPP}
\eea
which corresponds to quartic Horndeski with $G_4\,=\,1+\lambda \sqrt{X}$. 
A branch of spherically symmetric solutions is the Schwarzschild geometry written in Lemaitre coordinates as
\bea
\bar \pi&=& \tau
\\
d s^2&=&-d \tau^2+\frac{r_s}{r} d \rho^2+r^2\,\left(d \theta^2+\sin^2 \theta\,d \varphi^2\right)\,,
\eea
 where $r_s$ is the Schwarzschild radius and 
 \be
 \label{defofr1}
r\,=\,r_s^{1/3}\,\left[\frac32\,\left( \rho-\tau\right) \right]^{2/3}
 \,.\ee 
 Notice that the scalar profile depends on time only (and $\bar X\,=\,1$) hence its gradient is a Killing vector
 for the space-time we consider. The symmetry arguments of the main text
  prevent the propagation of scalar degrees of freedom. 
We now  show explicitly that the only dynamical degrees of freedom are the ones
of GR; on the other hand, their evolution equations are distinct from the ones of Einstein gravity. 
As anticipated, we use the methods and results of the recent paper \cite{Khoury:2020aya} which can be directly applied to our configuration,
and refer the reader to that work for the intermediate steps and for a clear discussion 
of the gauge conditions.
\smallskip

Linearised perturbations around spherically symmetric black holes are conveniently expanded in terms
of spherical harmonics $Y_{\ell m} (\theta, \varphi)$. The modes
 separate into quantities
that are {\it odd} or {\it even} under the parity symmetry $(\theta,\,\varphi)\,\to\,(\pi-\theta, \pi+\varphi)$. (Here $\pi$
denotes the number 3.1415..), that do not mix at the linearised level.  
We study them separately.

\subsection{Odd parity fluctuations}

This class of fluctuations only include metric perturbations, because scalar fluctuations are of even parity. 
 There are then
  two modes to be analysed (a third one can be 
gauged away using diffeomorphism invariance), usually denoted as 
$h_{(0)}^{\ell m}(\tau, \rho)$, $h_{(1)}^{\ell m}(\tau, \rho)$. The metric can be decomposed at linearised level as
(we focus on odd parity fluctuations with $\ell\ge2$)
\bea
ds^2&=&-d \tau^2+\frac{r_s}{r} d \rho^2+r^2\,\left(d \theta^2+\sin^2 \theta\,d \varphi^2\right)-
\nonumber
\\
&&-2\,d \tau\,d \theta\,\frac{1}{\sin \theta}\,\left[ \sum_{\ell=2}^\infty \sum_{m=-\ell}^\ell\,h_{(0)}^{\ell m}
\,\partial_\varphi Y_{\ell m}
\right]+2 \,d \tau\,d \varphi\,{\sin \theta}\,\left[ \sum_{\ell=2}^\infty \sum_{m=-\ell}^\ell\,h_{(0)}^{\ell m}
\,\partial_\theta Y_{\ell m}
\right]
\nonumber
\\
&&-2\,d \rho\,d \theta\,\frac{1}{\sin \theta}\,\left[ \sum_{\ell=2}^\infty \sum_{m=-\ell}^\ell\,h_{(1)}^{\ell m}
\,\partial_\varphi Y_{\ell m}
\right]+ 2\,d \rho\,d \varphi\,{\sin \theta}\,\left[ \sum_{\ell=2}^\infty \sum_{m=-\ell}^\ell\,h_{(1)}^{\ell m}
\,\partial_\theta Y_{\ell m}
\right]\,.
\eea
Plugging this decomposition in the action \eqref{actAPP}, one gets the effective Lagrangian  for the modes $h_{(0)}$, $h_{(1)}$ (from now on we understand the multipole indexes $(\ell, m)$ in the $h_{(i)}$)
\be \label{lagod1}
{\cal L}_{\rm odd}=\sqrt{\frac{r}{4 r_s}}\,\left(h_{(0)}'-\dot h_{(1)}- \sqrt{\frac{4\,r_s}{r^3}} 
(h_{(0)}+h_{(1)})\right)^2
 +\frac{\ell (\ell+1)-2}{2} \left(\sqrt{\frac{r_s}{r}} h_{(0)}^2-(1+\lambda) \sqrt{\frac{r}{r_s}} h_{(1)}^2 \right)^2\,,
\ee
with dot and prime denoting derivatives along $\tau$ and $\rho$ respectively. 
Notice that the scalar-tensor interactions
associated  with the additional coupling $\lambda$ in action \ref{actAPP} only controls the coefficient
depending on $\lambda$ in the last term of eq \eqref{lagod1}. The two quantities $h_{(0)}$ and $h_{(1)}$
are not independent: the corresponding field equations can be combined so to obtain a constraint between
the two fields, that can be solved by re-expressing the two quantities in terms of a single field $\Psi$ as follows:
\bea
h_{(0)}&=&\frac{1}{\ell (\ell+1)-2}\frac{r^2}{r_s} \left( r\,\Psi'+\frac52 \sqrt{\frac{r_s}{r}} \,\Psi\right)\,,
\\
h_{(1)}&=&\frac{1}{\ell (\ell+1)-2}\frac{r}{1+\lambda} \left( r\,\dot \Psi-\frac52 \sqrt{\frac{r_s}{r}} \,\Psi\right)\,.
\eea
 Plugging these expressions in eq \eqref{lagod1} one finds 
 
\bea\label{rewhe2}
{\cal L}_{\rm odd}\,=\,\frac{1}{2\left(1+\lambda\right)}\sqrt{\frac{r_s}{r}}\frac{r^2}{\ell (\ell+2)-2}
\left[ \dot \Psi^2-\frac{(1+\lambda) r}{r_s}\Psi'^2+\left(\frac{15\,r_s}{4 r^3}
+\frac{(1+\lambda)}{4 r^2}\left( 3-4 \ell (\ell+1)\right)\right)\Psi^2
\right]\,,\nonumber\\
\eea
 
 \noindent
 that we discussed as  eq \eqref{rewhe1} in the main text.

\subsection{Even parity fluctuations}

Scalar excitations can contribute to even parity perturbations, that we parameterize as 
\bea
\pi(x)&=&\tau+\sum_\ell \,\sum_{m=-\ell}^\ell \,\hat \pi^{\ell m}(\tau, \rho)\,Y_{\ell m} (\theta, \varphi)\,,
\\
h^{\rm even}_{\mu\nu}&=&\sum_\ell \,\sum_{m=-\ell}^\ell \,
\left(  
      \begin{array}{ccc}
 H_{(0)}^{\ell m}&  H_{(1)}^{\ell m}(\tau, \rho) & \alpha^{\ell m}\,\nabla_A \\
  H_{(1)}^{\ell m} &  \frac{r_s}{r}\,H_{(2)}^{\ell m}(\tau, \rho) & \beta^{\ell m}(\tau, \rho)\,\nabla_A \\
   \alpha^{\ell m}\,\nabla_A &
   \beta^{\ell m}\,\nabla_A
   & \left(r^2\,K^{\ell m}\,g_{AB}+Q^{\ell m}\,\nabla_A \nabla_B\right)
 \end{array} \right)
 \,Y_{\ell m}(\theta, \varphi)
 \nonumber\,,\\
\eea
respectively in the scalar and metric sector, and $g_{AB}$
parameterise the metric on the two-sphere (the capital Latin indexes denote coordinates on such
sphere). Understanding again the indexes $(\ell, m)$,  the quantities $\hat \pi$ (scalar fluctuation)
and $H_{(0)}$, $H_{(1)}$, $H_{(2)}$, $K$, $Q$, $\alpha$, $\beta$ (parity even metric fluctuations) depend
on $\tau$ and $\rho$.  Some of these quantities can be set to zero exploiting diffeomorphism invariance. 

 We shall now separate the discussion for modes with $\ell=0, 1$ and $\ell\ge 2$ with the specific aim to demonstrate  the absence of additional scalar modes with respect to GR, thanks to the 
  symmetry as discussed in the main text in section \ref{sec-gra}. 

  \subsubsection*{Monopole fluctuations, $\ell=0$}

For the monopole sector, the fluctuations $\alpha$, $\beta$, $Q$ are absent, while we can
set to zero $H_0$ and $K$ with a diffeomorphism transformation. The resulting effective Lagrangian
reads 
(recall that we use conventions where dot 
 and prime denote respectively derivatives along
$\tau$ and $\rho$)
\bea
{\cal L}_{\rm even}^{\ell=0}
&=&2\,\lambda\,\sqrt{\frac{r_s}{r}}\,\hat \pi'^2+\frac{1+\lambda}{2}\,\sqrt{\frac{r_s}{r}}\,H_2^2-2 \lambda\,
\sqrt{\frac{r_s}{r}}\,
H_2\,\hat \pi'+2 r\,H_1\,\dot H_2\,.
\eea
$H_1$ is a Lagrange multiplier that imposes the condition $H_2=0$ (up to a contribution
independent from time that can be re-adsorbed in the background). Using this
condition, we get
\bea
{\cal L}_{\rm even}^{\ell=0}
&=&2\,\lambda\,\sqrt{\frac{r_s}{r}}\,\hat \pi'^2\,.
\eea
Hence the monopole scalar fluctuation  does not propagate, since $\hat \pi$ only appears in the effective with spatial derivatives. 
This is a consequence of the symmetry
 arguments discussed in the main text.

  \subsubsection*{Dipole fluctuations, $\ell=1$}
In this case, a convenient gauge to choose is 
$$
H_0\,=\,\beta\,=\,r^2\,K-Q\,=\,0\,.
$$
The effective Lagrangian for $\ell=1$ fluctuations is 
\bea
{\cal L}_{\rm even}^{\ell=1}
&=&2\,\lambda\,\sqrt{\frac{r^3_s}{r^3}}\,
\left(
\frac{r}{r_s} \hat \pi'^2 -\frac{ \hat \pi^2}{r^2}\right)+
\frac{1+
\lambda}{2}  \sqrt{\frac{r_s}{r}}\,H_2^2
+2 \sqrt{\frac{r_s}{r}}\,H_2 \left[
\left(r \alpha \right)^{.}-
\lambda \,r\,\left(\frac{\hat \pi}{r}\right)'
\right]
\nonumber
\\
&&+\sqrt{\frac{r_s}{r}}\,H_1^2+2 H_1\,\left( r \dot H_2- \sqrt{\frac{r_s}{r}}\,\alpha'\right)+\sqrt{\frac{r_s}{r}}
\alpha'^2\,.
\eea
The field $H_1$ is non-dynamical, and is fixed by its equation of motion to
\be
H_1\,=\,-\sqrt{r_s r}\,\dot{H}_2+\alpha'\,.
\ee
Plugging this condition in the Lagrangian, one finds that the field $\alpha$ becomes a Lagrange multiplier, 
imposing the condition
\be\label{conH2a}
r\,\dot H'_2+2 \sqrt{\frac{r_s}{r}}\,\dot H_2+\frac{3 r_s}{2 r^2} H_2\,=\,0\,.
\ee
Using this information, the effective Lagrangian results

\bea
{\cal L}_{\rm even}^{\ell=1}
&=&2\,\lambda\,\sqrt{\frac{r^3_s}{r^3}}\,
\left(
\frac{r}{r_s} \hat \pi'^2 -\frac{ \hat \pi^2}{r^2}\right)+\lambda \sqrt{\frac{r_s}{r}}\,\hat \pi\,\left( 
2 H_2'+\sqrt{\frac{r_s}{r} }\,H_2
\right)
\nonumber
\\&&
-\sqrt{r_s r^3} \dot{H}_2^2+\sqrt{\frac{r_s}{4 r}}\,H_2^2\,,
\eea
where the field $H_2$ satisfies the constraint condition \eqref{conH2a}. The scalar field $\hat \pi$ is non-dynamical since it appears with no time derivatives in the effective Lagrangian, hence also monopole fluctuations do not rise to dynamical degrees of freedom.

  \subsubsection*{Multipole fluctuations wih $\ell\ge2$}

For higher multipoles the analysis is more involved, since in this case there
are propagating tensor degrees of freedom (as in Einstein gravity) whose dynamics
is influenced by the non-propagating scalar modes. It is hard to fully diagonalise
the equations of motion, on the other hand we shall prove that there are no additional
degrees of freedom with respect to GR.

A convenient gauge is
\be
H_0\,=\,K\,=\,Q\,=\,0\,.
\ee
We are left with five fields to analyse: $H_1$, $H_2$, $\alpha$, $\beta$, $\pi$. Exactly as for the monopole 
case $\ell=1$, the field $H_1$ is non-dynamical and is algebraically determined by its equations of motion. Plugging this information in the effective Lagrangian, one finds that also in this case $\alpha$ leads to a Lagrange multiplier that
imposes the condition
\bea
&&
r\,\dot H'_2+\left(\frac{\ell (\ell+1)}{2}+1 \right) \sqrt{\frac{r_s}{r}}\,\dot H_2+\frac{3 \,\ell\,(\ell+1)\,r_s}{4\, r^2} H_2\,=\,
\nonumber
\\
&=&
\frac{\ell (\ell+1)}{2} \left[
2 \sqrt{\frac{r}{r_s}}\,\dot \beta'
+\frac{3}{r}\,\dot \beta+\frac{\beta'}{r}+\frac{3}{r^2}\sqrt{\frac{r_s}{r}}\,\beta
\right]\,.
\label{lagcol2}
\eea
This Lagrange constraint  \eqref{lagcol2} is not easy to immediately interpret, but it says that the quantity 
$H_2$ is not an independent field. This can be made more manifest by the change of variable from $\beta$
to $\gamma$:
\be
\beta\,\equiv\,\gamma+\frac{\sqrt{r_s r}}{\ell (\ell+1)}\,H_2
\,,
\ee
which allows us to re-express the constraint  \eqref{lagcol2}  as 
\be \label{lagcol2a}
\partial_\tau\,\left[\frac{H_2}{r^{3/2}}\right]\,=\,\frac{3 \ell (\ell+1)}{\sqrt{r_s}\,\left(\ell^2+\ell-2\right)}
\left[ \frac{2}{3\sqrt{r_s}} 
\partial_\tau \left( \frac{\gamma'}{\sqrt{r}}\right)+\frac{1}{r} \partial_\tau \left( \frac{\gamma}{{r}}\right)
\right]\,.
\ee
Although we do not use the field $\gamma$ in what comes next, relation \eqref{lagcol2a} demonstrates
that the field $H_2$ can be expressed in terms of this quantity. (Notice that since $r$ depends on $\tau$, see
eq \eqref{defofr1}, it is nevertheless not immediate to integrate \eqref{lagcol2a}.)

\smallskip

Using the previous information,
the effective Lagrangian for the modes $\pi$, $H_2$, $\beta$  results
\bea
{\cal L}_{\rm even}^{\ell \ge 2}
&=&\frac{1-\ell-\ell^2}{2}\,\sqrt{\frac{r}{r_s}}\,\dot \beta^2
+2 r \dot H_2\,\dot \beta-\frac{2}{\ell (\ell+1)}\sqrt{\frac{r^4}{r_s}}\,\dot H_2^2
-2 \,\sqrt{\frac{r}{r_s}}\,\beta \,\dot H_2
+2 \,\lambda\,\sqrt{\frac{r}{r_s}}\,\pi'^2+\lambda\,\frac{2\,\ell (\ell+1)}{r}
\,\beta\,\hat \pi'
\nonumber
\\
&& \hskip-1.5cm -2\, \lambda\,\sqrt{\frac{r}{r_s}}\,H_2 \pi'-\frac{\ell (\ell+1)}{r^2}\,
\left(\sqrt{\frac{r}{r_s}}\,\beta^2
+r \,\beta H_2+2 \lambda \,\sqrt{\frac{r}{r_s}}\, \beta \,\hat \pi -2 \lambda\,r_s \,H_2  \,\hat\pi+ \lambda 
\sqrt{\frac{r^3}{r^3_s}}\,\hat \pi^2
\right)+\frac{1+\lambda}{2}\,\sqrt{\frac{r}{r_s}}\,H_2^2\,,
\nonumber
\\
&&
\label{effLagL2}
\eea
where the quantities $H_2$ and $\beta$ are related by the Lagrange constraint \eqref{lagcol2}, or alternatively
the relation \eqref{lagcol2a}, which shows that $H_2$ is not an independent field. Then Lagrangian \eqref{effLagL2} contains two independent modes, $\beta$ and $\pi$.  On the other hand, the scalar $\hat \pi$
does not appear with time derivatives, and   it is not difficult to realize that its equation of motion provides a constraint, algebraically
fixing $\pi$ in terms of the remaining quantities. This fact is more easily seen shifting $H_2$ to $K_2$ by the following
relation
\be
H_2\,\equiv\,K_2-2\,\ell (\ell+1)\,\sqrt{\frac{r_s}{r^3}}\,\pi+2\, \pi'\,.
\ee
When $\lambda\neq0$, 
the equation of motion for $\pi$ provides the following constraint, when expressed
in terms of $H_2$, $\beta$:
\be
\ell \,(\ell+1)\,(2 \ell^2+2 \ell-r)\,\pi\,=\,\sqrt{\frac{r^3}{4 r_s}}\left(2 \ell (\ell+1)-1 \right)\,K_2
+\frac{r^3}{r_s}\,K_2'+\sqrt{\frac{r^5}{r_s^3}}\,\ell (\ell+1)\,\beta'\,.
\ee
This relation -- as well as \eqref{lagcol2a} -- demonstrate that the effective Lagrangian \eqref{effLagL2} describes
the dynamics of a single propagating degree of freedom, as in  Einstein gravity, and our scalar-tensor
interactions do not excite additional modes around our black hole configuration, as expected by our 
symmetry arguments. 

\end{appendix}

\mciteSetMidEndSepPunct{}{\ifmciteBstWouldAddEndPunct.\else\fi}{\relax}

\providecommand{\href}[2]{#2}\begingroup\raggedright\endgroup

\end{document}